\documentclass{jkas}

\def\year{2025} % publication year
\def\volume{56} % journal volume
\def\issue{1} % journal issue
\def\beginpage{1} % first page of article
\def\received{---} % date paper was received by JKAS
\def\accepted{---} % date of acceptance
\def\published{---} % date of publication
\date{Received \received; Accepted \accepted; Published \published}

\title{%
Cosmological Distance Measurements of High Redshift Blazars: OJ~248 (z~=~0.939) and 4C~+38.41 (z~=~1.814)
}

\author[1,2]{Young-Bin~Shin}{0009-0006-9248-0022}
\author[1,2,$\star$]{Sang-Sung~Lee}{0000-0002-6269-594X}
\author[2]{Sincheol~Kang}{0000-0002-0112-4836}
\author[2]{Whee~Yeon~Cheong}{0009-0002-1871-5824}
\author[1,2]{Chanwoo~Song}{0009-0003-8767-7080}
\author[3]{Jeffrey~A.~Hodgson}{0000-0001-6094-9291}
\affil[1]{University of Science and Technology, Gajeong-ro 217, Yuseong-gu, Daejeon 34113, Republic of Korea}
\affil[2]{Korea Astronomy and Space Science Institute, Daedeok-daero 776, Yuseong-gu, Daejeon 34055, Republic of Korea}
\affil[3]{Department of Physics and Astronomy, Sejong University, 209 Neungdong-ro, Gwangjin-gu, Seoul 05006, Republic of Korea}
\def\runningauthor{%
Shin et al.
}

\def\runningtitle{%
Cosmological Distance Measurements of High Redshift Blazars: OJ~248~($z=0.939$) and 4C~+38.41~($z=1.814$)
}

\def\keywords{%
methods: observational – techniques : interferometric – cosmology : observations – cosmology : distance scale – radio continuum : galaxies.
}

\def\abstracttext{%
In this study, we estimated the angular diameter distances to two high-redshift active galactic nuclei~(AGNs), OJ~248~($z=0.939$) and 4C~+38.41~($z=1.814$), using 43~GHz radio light curves.
The aim of this work is to extend AGN variability–based distance measurement methods to the high-redshift regime.
The distance estimates were analyzed under two assumptions for the maximum intrinsic brightness temperature $T_{\text{B,~int}}$:~(1)~the equipartition temperature, and (2)~the sample-averaged $T_{\text{B,~int}}$.
As a representative result, when adopting the equipartition temperature, the angular diameter distances to OJ~248 and 4C~+38.41 are estimated to be $7592.7\pm396.7$~Mpc and $11069.8\pm1216.9$~Mpc, respectively.
In addition, Doppler factors were calculated using the inverse-Compton method, and additional distance estimates were obtained based on these values.
Our error budget analysis shows that the most significant systematic uncertainty arises from epoch selection.
These results indicate that systematic effects have a substantial impact on the derived distance estimates and limit their reliability, particularly in the high-redshift regime.
Reducing these uncertainties will require improved observational cadence and reduced post-fit noise.
AGN variability provides an alternative approach to distance estimation.
However, our results reveal significant limitations in the current methodology and indicate that further methodological and observational improvements are required before its cosmological applicability can be reliably assessed.
}

\usepackage{threeparttable}
\providecommand{\corrauthor}{}
\begin{document}
\jkashead %% set title, authors, abstract, etc.

%%%%%%%%%%%%%%%%%%%%%%%%%%%%%%%%%%%%%%%%%%%%%%%%%%%%%%%%%%%%%%%%%%%%%
%%% BEGIN MAIN TEXT HERE %%%%%%%%%%%%%%%%%%%%%%%%%%%%%%%%%%%%%%%%%%%%
%%%%%%%%%%%%%%%%%%%%%%%%%%%%%%%%%%%%%%%%%%%%%%%%%%%%%%%%%%%%%%%%%%%%%
\section{Introduction}
\label{Introduction}
A standard candle is a celestial body whose intrinsic brightness~(absolute magnitude) is known.
It serves as a fundamental tool for measuring cosmological distances.
Representative examples of standard candles include Cepheid variables and Type Ia supernovae.
Through observations of Cepheid variable stars in the Small Magellanic Cloud, \citet{Leavitt1912} found that the logarithm of the period is linearly related to the logarithm of the star's average intrinsic optical luminosity.
Through this relationship, the luminosity of the Cepheid variable star was determined, and the distance to the Small Magellanic Cloud was measured using the inverse square law.
The current limit of the distance that can be measured with Cepheid variables is about 20.4$\pm$1.7~Mpc, which is the distance to M100~\citep[$z=0.005$,][]{Allison2014} in the Virgo cluster~\citep{Narasimha1998}.
\citet{Hamuy1993} later discovered Type Ia supernovae through the Cal$\acute{a}$n/Tololo Supernova Survey and began using them as standard candles.
Because Type Ia supernovae explode with nearly identical masses~($M_{\text{chan}}=1.4M_{\odot}$), they exhibit a consistent absolute magnitude~($M_{\text{V}}=-19.3\pm0.03+5\log\left(\frac{H_{0}}{60}\right)$), which enables reliable distance estimation~\citep{Hillebrandt2000}.
In particular, \citet{Brout2022} presented the Pantheon+ compilation of over 1,700 Type Ia supernovae, in which the luminosity of each supernova was standardized through the improved light curve modeling~(SALT2-B22) and comprehensive bias corrections.
These refinements allowed the intrinsic dispersion of standardized magnitudes to be minimized, reinforcing the role of Type Ia supernovae as highly precise cosmological distance ladder.
The most distant Type Ia supernova is GND12Col~($z = 2.26$)~\citep{Scolnic2018, Riess2018}.
Standard candles were used to develop and improve the cosmological model~(e.g.,~$\Lambda$CDM~($\Lambda$ Cold Dark Matter)) and to determine the Hubble constant $H_{0}$.
In particular, distant Type Ia supernovae~(e.g.,~SN1997ck~($z=0.97$)) were used to reveal that the universe is expanding at an accelerated rate, and Cepheid variables and Type Ia supernovae played an important role in constraining the value of the Hubble constant~\citep{Riess1998, Perlmutter1999}.
Representative recent estimates of the Hubble constant come from observations of the cosmic microwave background~(CMB) by \citet{Planck2020}~($H_{0}=67.40\pm0.50$~km/s$\cdot$Mpc) and from observations of Cepheid variables by \citet{Riess2022}~($H_{0}=73.04\pm1.04$~km/s$\cdot$Mpc).

\citet{Hodgson2020} measured the distance to the active galactic nuclei~(AGN) 3C~84~($z=0.0178$) by analyzing variations in its radio luminosity during flaring events.
They subsequently proposed AGNs as potential new cosmological distance ladders~\citep{Hodgson2023}.
Building on these studies, we estimate the angular diameter distances to two high-redshift blazars---OJ~248~($z=0.939$) and 4C~+38.41~($z=1.814$)---and compare them with distances expected from the $\Lambda$CDM model~($\Omega_{m}=0.315$, $\Omega_{\Lambda}=0.685$; Table~\ref{tab1}), to examine the characteristics of the distance estimates obtained with this method.

\begin{table*}[h]
    \centering
    \begin{threeparttable}
        \caption{Basic information of sources (OJ~248 and 4C~+38.41)}
        \label{tab1}
            \begin{tabular}{cccccc}
                \toprule[1.5pt]
                Name & Category & R.A.~(J2000) & Dec~(J2000) & $z$ & $D_{\text{A}}$~(Mpc) \\
                (1) & (2) & (3) & (4) & (5) & (6) \\ \hline
                OJ 248 & Blazar & 8h 30m 52.086s & +24$^\circ$ 10$^\prime$ 59.82$^{\prime\prime}$ & 0.939 & 1607.2$\pm$79.5 \\ \hline
                4C +38.41 & Blazar & 16h 35m 15.493s & +38$^\circ$ 8$^\prime$ 4.501$^{\prime\prime}$ &  1.814 & 1714.4$\pm$84.7 \\
                \bottomrule[1.5pt]
            \end{tabular}
        \begin{tablenotes}
	   \small
            \item{Note.~(1)~common name, (2)~source category obtained from \citet{Jorstad2017}, (3)~Right Ascension, (4)~Declination, (5)~redshift, (6)~angular diameter distance calculated according to $H_{0}=70.22\pm3.59$~km/s$\cdot$Mpc~\citep{Wright2006}.}
         \end{tablenotes}
    \end{threeparttable}
\end{table*}

\section{Observational data and analysis}
\label{Observational_data_and_analysis}
The radio observation data were obtained from Boston University's blazar monitoring (VLBA-BU-BLAZAR) program\footnote{https://www.bu.edu/blazars/BEAM-ME.html}~\citep{Jorstad2017, Weaver2022}, which monitors AGNs at 43~GHz.
We used data from November~2009 to February~2014 for OJ~248, and from December~2010 to February~2019 for 4C~+38.41, as these periods coincide with the detection of gamma-ray flares from both sources by the Fermi-LAT satellite.
The X-ray data used in this paper were obtained from the Open Universe for Blazars program\footnote{https://openuniverse.asi.it/blazars/swift/}~\citep{Giommi2021}.
This program frequently observed blazars using the Swift observatory, and in this study, the flux densities at an energy of 1~keV were used.

We measured the angular sizes and flux densities of the blazars by fitting circular Gaussian models to individual components in the images of the blazar monitoring observations.
The standard routines of the program DIFMAP~\citep{Shepherd1997} were used when fitting the model.

Fitting a circular Gaussian model yields parameters such as the total flux density, the location, and the size of each component.
The fitting parameters of components in VLBI images of radio sources are $S_{\text{tot}}$~(total flux density), $S_{\text{peak}}$~(peak flux density), $\sigma_{\text{rms}}$~(post-fit rms), and $d$~(Full Width Half Maximum, hereafter FWHM).
\citet{Fomalont1999} suggested the uncertainty of the fit parameters as follows:
\begin{align*}
    \sigma_{\text{peak}}=\sigma_{\text{rms}}\left(1+\frac{S_{\text{peak}}}{\sigma_{\text{rms}}}\right)^{\frac{1}{2}}, \\
    \sigma_{\text{tot}}=\sigma_{\text{peak}}\left(1+\frac{S_{\text{tot}}^{2}}{S_{\text{peak}}^{2}}\right)^{\frac{1}{2}}, \\
    \sigma_{\text{d}}=d\frac{\sigma_{\text{peak}}}{S_{\text{peak}}}, \tag{1}
    \label{eq1}
\end{align*}
where $\sigma_{\text{tot}}$, $\sigma_{\text{peak}}$, and $\sigma_{\text{d}}$ are the total flux density, the peak flux density, and the FWHM uncertainty of the components, respectively.
When determining the FWHM $d$ of a component, the resolution limit~\citep{Lobanov2005} must be taken into account, where the minimum resolvable size of the component in the image is
\begin{align}
    d_{\text{min}}=\frac{2^{\frac{1+\beta}{2}}}{\pi}\left(\pi a b \ln 2 \ln \frac{\left(S/N\right)}{\left(S/N\right)-1}\right)^{\frac{1}{2}}, \tag{2}
    \label{eq2}
\end{align}
where $a$ and $b$ are the axes of the restoring beam, S/N is the signal-to-noise ratio, and $\beta$ is a weighting function, 0 for natural weighting and 2 for uniform weighting.
In this study, we obtained the upper limit of the component sizes as $d_{\text{min}}$ when $d<d_{\text{min}}$.
The model fit parameters are tabulated in Appendix \ref{Table_of_the_circular_Gaussian_model-fitted_values}.

\section{Results}
\label{Results}
\begin{figure*}[h!]
    \centering
    \includegraphics[width=\linewidth]{Integrated_graph.png}
    \caption{Results of distance determination for OJ~248~(left panels) and 4C~+38.41~(right panels) \\ Note.~(a),(e):~light curve of core, (b),(f)~: major FWHM of core, (c),(g)~: inverse-Compton Doppler factor, (d),(h)~: estimated distances. The blue dots in the first row are radio flux density, and the black dots are X-ray flux density. Additionally, the solid red line is a simple linear regression line for the change in logarithmic flux density~($\ln{(\text{Flux density})}$) over time in the range from the starting point to the peak. The red dotted lines in the last rows represent the distances obtained from the $\Lambda$CDM model. The gray area represents the area defined as flare in this study.}
    \label{fig1}
\end{figure*}

\subsection{Calculating variability time-scale $\Delta t$}
\label{Calculating_variability_time-scale}
We defined a flare as an event in which the core radio flux density of OJ~248 and 4C~+38.41 increased from a local minimum to a maximum.
A flare was identified only if the radio flux density increased over at least three consecutive observational epochs.
The peak of each flare was defined as the local maximum that appeared after the corresponding flaring segment, which was selected as the point with the visually distinct maximum flux density on the light curve.
For the subsequent distance calculations, we use the flux density and angular size measured at the peak of each flare.
This choice is physically motivated by the light-crossing-time argument, which requires an emitting region~(e.g., a core component) to collide with a moving shock in the jet and to peak in flux density with being resolved from the moving shock, in order to provide a meaningful estimate of its intrinsic size.
Therefore, adopting peak-epoch parameters ensures consistency across all flares and provides a practical reference point for estimating the properties of the variable emission region.

A simple linear regression analysis was performed on the change in logarithmic flux density~($\ln{(\text{Flux density})}$) over time in the range of the starting point and peak point.
This is because flaring in blazar jets can generally be described by an exponential flux increase $S_{\nu}\left(t\right)\propto e^{kt}$~\citep{Terasranta1994, Valtaoja1999, Jorstad2005, Hovatta2009, Jorstad2017}.
Applying a logarithmic transformation transforms this exponential fluctuation into a linear relationship over time, allowing a straight-line fit to accurately capture the upward trend and derive the slope $k$.
The variability time-scale of a radio core component is obtained as $\Delta t=1/k$~\citep{Hodgson2020}, and the results are summarized in Table~\ref{tab2}.

\begin{table*}[h]
    \centering
    \caption{Identified flares parameters}
    \begin{tabular}{ccccc}
        \toprule[1.5pt]
        Source & \begin{tabular}[c]{@{}c@{}}Flare\\No.\end{tabular} & \begin{tabular}[c]{@{}c@{}}Start Epoch\\(MJD)\end{tabular} & \begin{tabular}[c]{@{}c@{}}End Epoch\\(MJD)\end{tabular} & $\Delta t$ (day) \\ \hline
        OJ~248 & 1 & 55163 & 55335 & 138.8$\pm$11.5 \\
         & 2 & 55361 & 55820 & 87.7$\pm$3.0 \\
         & 3 & 56282 & 56713 & 53.9$\pm$2.1 \\ \hline
        4C~+38.41 & 1 & 55534 & 55820 & 375.8$\pm$62.7 \\
         & 2 & 55991 & 56282 & 206.7$\pm$9.6 \\
         & 3 & 56307 & 57020 & 144.0$\pm$5.9 \\
         & 4 & 58110 & 58187 & 134.2$\pm$22.1 \\
         & 5 & 58227 & 58522 & 122.2$\pm$12.3 \\
        \bottomrule[1.5pt]
    \end{tabular}
    \label{tab2}
\end{table*}

\subsection{Estimating $D_{\text{A}}$ using maximum intrinsic brightness temperature}
\label{Estimating_distance_using_temperature}
Following \citet{Hodgson2023} and \citet{Cheong2025}, we calibrated relativistic effects by assuming that AGNs reach a maximum intrinsic brightness temperature.

The angular diameter distance~$D_{\text{A}}$ assuming that $c\Delta t$ relates to the radius of the emission region is:
\begin{align}
    D_{\text{A}} = X\frac{2 \ln{2} c^{3} S \Delta t}{\pi k_{\text{B}} T_{\text{B,~int}} \nu^{2} \theta_{\text{FWHM}}^{3}}, \tag{3}
    \label{eq3}
\end{align}
where $c$ is the speed of light~(cm/s), $S$ is the flux density~(Jy) in the emission region, $\Delta t$ is the variability time-scale~(s) and $X$ is a scaling factor for geometric assumptions, such that $X\simeq0.70$ for a disk and $X\simeq0.74$ for a sphere.
Additionally, $k_{\text{B}}$ is the Boltzmann constant~(erg/K), $T_\text{B,~int}$ is the intrinsic brightness temperature~(K) of the emission region, $\nu$ is the observing frequency~(Hz), and $\theta_{\text{FWHM}}$ is the measured angular size of the emission region~(i.e.,~the FWHM of a circular Gaussian model fitted to the emission region, (mas)).
In Equation~(\ref{eq3}), the value of $S$ corresponds to the peak flux density during the flare.

When estimating the angular diameter distance~$D_{\text{A}}$ of OJ~248 and 4C~+38.41, the following two cases were considered for the value of $T_{\text{B,~int}}$:~(1)~equipartition temperature~$T_{\text{eq}}$, and (2)~average value from samples of multiple sources $T^{\text{Cheong}}_{\text{B,~int}}$.
The results of calculating the distance using this are presented in~\ref{Estimated_distances_from_each_source}.
In the following subsections, we report as representative values the mean angular diameter distances of the two sources, derived under different assumptions for $T_{\text{B,~int}}$ using the peak-epoch parameters defined in Section~\ref{Calculating_variability_time-scale}, namely the flux density, angular size, and variability time-scale.

\subsubsection{Case 1:~Equipartition temperature~$T_{\text{eq}}$}
\label{Case_1}
\cite{Readhead1994} proposed an intrinsic brightness temperature of $T_{\text{eq}}=5\times10^{10}$~K, assuming equipartition between particle and magnetic field energy densities.
The representative angular diameter distances of the two sources in this case are estimated to be $7592.7\pm396.7$~Mpc for OJ~248 and $11069.8\pm1216.9$~Mpc for 4C~+38.41, as shown in Figure~\ref{fig1}d,~h.

\subsubsection{Case 2:~Average value from samples of multiple sources~$T^{\text{Cheong}}_{\text{B,~int}}$}
\label{Case_2}
\citet{Cheong2025} estimated the intrinsic brightness temperature $T_{\text{B,~int}}$ for individual blazars in their sample based on variability analysis and reported an average value across all sources (Table~\ref{tab4}).
We applied the 43~GHz average to estimate the angular diameter distances for OJ~248 and 4C~+38.41, as shown in Figure~\ref{fig1}d,h.
When $T^{\text{Cheong}}_{\text{B,~int}}$ is applied, the representative value of the estimated angular diameter distance shows a relatively large dispersion compared to {\textbf{the equipartition case.}}

\begin{table}[h]
    \centering
    \begin{threeparttable}
        \caption{Frequency dependent constraints on $T_{\text{B,~int}}$}
        \label{tab4}
            \begin{tabular}{cc}
                \toprule[1.5pt]
                Freq. & $\log_{10}T_{\text{int}}$ \\
                (GHz) & $H_{0}=67.40\pm0.50$~km/s$\cdot$Mpc \\ \hline
                15 & $10.13<\log_{10}T_{\text{int}}<11.56$ \\
                24 & $9.73<\log_{10}T_{\text{int}}$ \\
                43 & $9.12<\log_{10}T_{\text{int}}<11.65$ \\
                86 & $9.29<\log_{10}T_{\text{int}}$ \\
                \bottomrule[1.5pt]
            \end{tabular}
        \begin{tablenotes}
	   \small
            \item{Note.~The upper limit is derived from source variability, and the lower limit is derived from population analysis. For all values, the core geometry is assumed to be a uniform disk.}
         \end{tablenotes}
    \end{threeparttable}
\end{table}

\subsection{Calculating Doppler factor $\delta$ and estimating $D_{\text{A}}$ using $\delta$}
\label{Calculating_doppler_factor}
Estimating distance via Doppler boosting requires prior determination of the Doppler factor $\delta$.
Among various available methods, we adopted the inverse-Compton method, which assumes that the dominant emission mechanism at X-ray frequencies is the synchrotron self-Compton~(SSC).
Supporting this assumption, previous spectral energy distribution~(SED) studies~\citep{Carnerero2015, Zhou2024} confirm that SSC emission occurred in these sources during the relevant epochs.
Following \citet{Liodakis2017}, the Doppler factor $\delta_{\text{IC}}$ is computed under the assumption of a uniform magnetic field and a power-law electron energy distribution, as follows:
\begin{align}
    \delta_{\text{IC}} = f\left(\alpha\right) F_{\text{m}} \left(\frac{\ln{\left(\frac{\nu_{\text{b}}}{\nu_{\text{m}}}\right)}}{F_{\chi} \theta_{\text{VLBI}}^{6+4\alpha} \nu_{\chi}^{\alpha} \nu_{\text{m}}^{5+3\alpha}}\right)^{\frac{1}{4+2\alpha}}\left(1+z\right), \tag{4}
    \label{eq4}
\end{align}
where $z$ is the redshift and $\nu_{\text{b}}$ is the synchrotron high-energy cutoff, estimated to be $10^{14}$~Hz.
$\nu_{\text{m}}$ is the radio observation frequency~(GHz) and $F_{\text{m}}$ is the synchrotron flux density~(Jy) of the emission region at frequency $\nu_{\text{m}}$.
Additionally, $\theta_{\text{VLBI}}$ is the angular size~(mas) of the corrected circular Gaussian model of the emission region, $\nu_{\chi}$ is the frequency corresponding to the X-ray observation photon energy of 1~keV, and $F_{\chi}$ is the X-ray flux density~(Jy).
The X-ray flux density $F_{\chi}$ was taken from the epoch closest in time to the corresponding radio observations.
The function $f\left(\alpha\right)$ is given by $f\left(\alpha\right) \simeq 0.08\alpha+0.14$, where $\alpha$ is the optically thin spectral index~\citep{Ghisellini1987} assumed to be $\alpha = 0.75$.
The resulting Doppler factors are shown in Figure~\ref{fig1}c,i, ranging from 0.27$\pm$0.04 to 14.8$\pm$0.8 for OJ~248 and from 6.0$\pm$0.1 to 89.0$\pm$8.1 for 4C~+38.41.

\citet{Cheong2025} derived a method based on the assumption that the variability time scale $\Delta t$ corresponds to the physical size of the emission region via the relation $a=fc\Delta t$, leading to the following expression for the angular diameter distance:
\begin{align}
    D_{\text{A}} = \frac{fc\Delta t\delta}{\theta_{\text{VLBI}} \left(1+z\right)}, \tag{5}
    \label{eq5}
\end{align}
where $c$ is the speed of light~(m/s) and $\Delta t$ is the variability time-scale~(s) obtained in Section \ref{Calculating_variability_time-scale}.
The Doppler factor $\delta$ is the inverse-Compton Doppler factor estimated following Equation~(\ref{eq4}).
Additionally, $z$ is the redshift and $f$ is a correction factor that equates the causality size to the physical linear size.
To ensure consistency with the distance estimates obtained using Equation~(\ref{eq3}), we adopt $f=1.0$ in Equation~(\ref{eq5}), such that the variability time-scale $\Delta t$ traces the diameter of the variable emission region.
For direct comparison with the distances obtained in Section~\ref{Estimating_distance_using_temperature}, we evaluate Equation~(\ref{eq5}) using the same peak-epoch parameters defined in Section~\ref{Calculating_variability_time-scale}; specifically, $\Delta t$, $\theta_{\text{VLBI}}$, and $\delta_{\text{IC}}$ are all taken at the flare peak-epoch.
Using this assumption, we estimate the angular diameter distance for individual flares and report the mean value as a representative distance for each source, following the same convention adopted in Section~\ref{Estimating_distance_using_temperature}.
The resulting mean angular diameter distances are $1205.7\pm81.4$~Mpc for OJ~248 and $3434.9\pm431.3$~Mpc for 4C~+38.41.

\section{Discussion}
\label{Discussion}
\subsection{Optimal parameter sets for distance measurements}
\label{Optimal_parameter_sets}
As described in the previous sections, We estimated the cosmological distances of the two target sources using Equation~(\ref{eq3}), which depends on four key parameters: the flux density $S$, the variability time-scale $\Delta t$, the angular size of the emission region $\theta_{\text{FWHM}}$, and the intrinsic brightness temperature $T_{\text{B,~int}}$.
We computed distances using a set of parameters, including $S$ and $\theta_{\text{FWHM}}$, obtained at the peak of the flare based on two intrinsic brightness temperature assumptions, and compared these results with the distance range obtained from the $\Lambda$CDM model.

Looking at the light curves of OJ~248 and 4C~+38.41~(Figure~\ref{fig1}a,e), flares~2 and 3 at OJ~248, and flares~1, 3, and 5 at 4C~+38.41 started with relatively high flux densities.
Therefore, they are not well-decomposed single flares.
Overlapping flares affect both $S$ and $\Delta t$ in Equation~(\ref{eq3}), likely leading to an overestimation of the distance.

To improve the accuracy of distance estimation, identifying well-isolated radio flares is essential, and higher-cadence observations are required.
The radio observation data used in this study have an average cadence of about a month, leading to a peaking time offset of half a month, and to 8$\%$ underestimated flux density measurement for an average variability time-scale of 157.9~days.
To better estimate the peak flux density $S$ of the variable emission region with an uncertainty of $<5\%$, it is necessary to improve the observation cadence to biweekly.

\subsection{Error budget analysis in $D_{\text{A}}$ determination}
\label{Error_budget_analysis}
In this section, we use the data summarized in Table~\ref{tabA1}-\ref{tabA2} and \ref{tabB1}-\ref{tabB2} to evaluate the statistical effect of four parameters on the overall uncertainty of $D_{\text{A}}$: the flux density~$S$, the variability time-scale~$\Delta t$, and the angular size of the emission region~$\theta_{\text{FWHM}}$, with the results shown in Table~\ref{tab6}.

\begin{table}[h]
    \centering
    \caption{Statistical error budget of each parameter}
    \begin{tabular}{ccc}
        \toprule[1.5pt]
        Parameter & OJ~248 & 4C~+38.41 \\ \hline
        $S$ & 0.7---10.3$\%$ & 0.9---11.6$\%$ \\
        $\Delta t$ & 3.4---8.3$\%$ & 4.1---16.7$\%$ \\
        $\theta_{\text{FWHM}}$ & 0.0---6.8$\%$ & 0.6---8.4$\%$ \\ \hline
        $D_{\text{A}}$ & 33.2---40.6$\%$ & 32.9---44.4$\%$ \\
        \bottomrule[1.5pt]
    \end{tabular}
    \label{tab6}
\end{table}

The uncertainties of the flux density $S$, variability time-scale $\Delta t$, and emission region $\theta_{\text{FWHM}}$ are $\sigma_{\text{tot}}$, $\frac{\sigma_{\text{tot}}}{S}$, and $\sigma_{\text{d}}$, respectively, (see Eq.\ref{eq1}) and these values are related to $\sigma_{\text{rms}}$.
Therefore, reducing $\sigma_{\text{rms}}$ would directly decrease the uncertainties in $S$, $\Delta t$, and $\theta_{\text{FWHM}}$, thereby improving the accuracy of $D_{\text{A}}$.
$\sigma_{\text{rms}}$ is proportional to the theoretical thermal noise $\Delta I_{\text{m}}$, which is calculated as follows:
\begin{align}
    \Delta I_{\text{m}} = \frac{\text{SEFD}}{\eta_{\text{c}}\sqrt{n_{\text{pol}}N(N-1)t_{\text{int}}\Delta\nu}}, \tag{6}
    \label{eq6}
\end{align}
where SEFD is the system equivalent flux density~(Jy), $\eta_{\text{c}}$ is the correlator efficiency, $n_{\text{pol}}$ is the number of polarization products included in the image, $N$ is the number of antennas, $t_{\text{int}}$ is the total on-source integration time~(s), and $\Delta\nu$ is bandwidth~(Hz). 
To reduce the theoretical thermal noise $\Delta I_{\text{m}}$, either (1)~increasing $N$, (2)~increasing $t_{\text{int}}$, or (3)~widening $\Delta\nu$ are required.
For example, halving $\Delta I_{\text{m}}$ would require either doubling $N$, increasing $t_{\text{int}}$ by a factor of four, or expanding $\Delta\nu$ by a factor of four.
If $\Delta I_{\text{m}}$ is halved, $\sigma_{\text{rms}}$ will also be halved, resulting in a halving of the uncertainties in $S$, $\Delta t$, and $\theta_{\text{FWHM}}$.
Consequently, the uncertainty in $D_{\text{A}}$ can be reduced by up to 12$\%$~(e.g., from 33.2$\%$ to 29.2$\%$ for OJ~248, from 32.9$\%$ to 29.0$\%$ for 4C~+38.41).

In addition to statistical uncertainties, our distance estimates are subject to several systematic effects associated with the underlying assumptions of the method.
First, the analysis relies on the assumption that the light-travel size $c\Delta t$ represents the physical size of the VLBI emission region.
In this work, this connection is interpreted on the basis of a causality argument:~intrinsic variations cannot occur on time-scales shorter than the light-crossing time of the emitting region~(modulo relativistic effects).
Therefore, $\Delta t$ provides a physically motivated upper limit on the characteristic size of the emitting region, rather than implying a strict one-to-one equivalence with the light-crossing time.
Considering the geometric scaling factors discussed in shock-in-jet models~(e.g., \citet{Marscher1985, Jorstad2005}), additional uncertainties may arise in relating the variability time-scale to the physical size of the emission region, and the magnitude of this uncertainty is difficult to quantify precisely.

Second, the derived angular diameter distance $D_{\text{A}}$ is highly sensitive to the choice of reference epoch and to the temporal evolution of the core structure.
Recalculations using the flare starting epoch and the minimum core-size epoch, in addition to the peak epoch, show that both flux density and $\theta_{\text{FWHM}}$ vary substantially, leading to significant changes in $D_{\text{A}}$.
Furthermore, by examining changes in $\theta_{\text{FWHM}}$ during the flare over half of the typical observing interval, corresponding to approximately two weeks, we find variations of $\theta_{\text{FWHM}}$ up to 25$\%$ for OJ~248 and 10$\%$ for 4C~+38.41.
These may yield potential changes of $D_{\text{A}}$ by the factors of 0.5-2.4 for OJ~248 and 0.8-1.4 for 4C~+38.41~(following $\frac{D_{\text{A}}\pm\Delta D_{\text{A}}}{D_{\text{A}}}\sim\left(1\mp\frac{\Delta\theta_{\text{FWHM}}}{\theta_{\text{FWHM}}}\right)^{-3}$).
Shifting the adopted peak epoch by $\pm1$~epoch results in fractional changes of 40–45$\%$ in $\Delta t$, which propagate linearly into the distance estimates.
These tests demonstrate that epoch selection and temporal structural evolution constitute the dominant sources of systematic uncertainty in the present analysis.

Finally, the VLBI core is modeled as a simple circular Gaussian component.
Although ultra-high-resolution observations have revealed structural complexity in some nearby sources, available high-frequency VLBI data indicate that, at the angular resolution relevant to this study, the core regions of our targets are adequately described by a single Gaussian component.
We therefore do not expect this modeling approximation to introduce a dominant systematic effect in the derived distances.

In summary, while statistical uncertainty can be partially mitigated by improving observation sensitivity and cadence, the precision of angular distance estimates are currently limited by systematic effects.
In particular, systematic uncertainties related to epoch selection and the interpretation of variability time-scales can be comparable to or greater than the statistical error budget and must be considered when assessing the reliability and precision of this methodology.

\subsection{Limitations and prospects of distance estimation at high redshift}
\label{Limitations_and_prospects}
In this study, we present the average of the angular diameter distances between two sources estimated from individual flares using the peak-epoch parameters defined in Section~\ref{Calculating_variability_time-scale} under different $T_{\text{B,~int}}$ assumptions as representative values.
These representative values show substantial systematic uncertainties depending on the adopted intrinsic brightness temperature assumptions and observational conditions.
These issues can be discussed from two perspectives:~(1)~the choice and applicability of the adopted $T_{\text{B,~int}}$ assumptions, and (2)~observational challenges in high-redshift blazars.

\subsubsection{Choice and applicability of $T_{\text{B,~int}}$ assumptions}
\label{Choice_and_applicability}
The $T^{\text{Cheong}}_{\text{B,~int}}$ assumption, presented in Section~\ref{Case_2}, shows a much larger dispersion of the estimated angular distances compared to {\textbf{the equipartition assumption.}}
This wide distribution is mainly due to the large uncertainty in the variability time-scale used to derive $T_{\text{B,~int}}$.
\cite{Cheong2025} estimated the variability time-scale based on observations of consecutive epochs.
Applying the e-folding time-scale to multiple epochs of flares, as performed in this study, or applying the flare decomposition technique proposed by Kang et al. (in prep), is expected to reduce this uncertainty and more strictly constrain the intrinsic brightness temperature.
This will ultimately lead to improved accuracy in distance estimation.

\subsubsection{Methodological limitations and future observational requirements}
\label{Methodological_limitations}
In the present analysis, systematic uncertainties are strongly affected by the limited observational cadence of the radio monitoring data.
Since the 43~GHz VLBI observations used in this study have an average cadence of approximately one month, significant uncertainties remain in determining the flare peak epoch and variability time-scale.
These effects directly propagate into the angular diameter distance estimates and limit the precision of the current methodology.

Furthermore, higher-cadence monitoring data for high-redshift blazars such as OJ~248 and 4C~+38.41 are absent.
Compared to nearby AGNs, long-term VLBI monitoring programs for high-redshift sources are relatively few, and densely sampled observations are not readily available.
As a result, reducing the current level of systematic uncertainty is difficult with the presently available datasets.

Therefore, the current variability-based distance estimation framework cannot yet be considered robustly applicable to high-redshift blazars.
Future improvements will require dedicated high-cadence monitoring observations, more rigorous flare decomposition techniques, and improved constraints on the physical relation between variability time-scales and the size of the emitting region.

\section{Conclusion}
\label{Conclusion}
In this study, we analyzed 43~GHz radio flare light curves to estimate the angular diameter distances to two high-redshift blazars, OJ~248 and 4C~+38.41, thereby extending AGN variability–based distance measurement methods to the high-redshift regime.
The distance estimates were obtained under different assumptions for the intrinsic brightness temperature.

As a representative result, when adopting the equipartition temperature, the angular diameter distances to OJ~248 and 4C~+38.41 are estimated to be $7592.7\pm396.7$~Mpc and $11069.8\pm1216.9$~Mpc, respectively.
In addition, Doppler factors derived using the inverse-Compton method range from 0.27$\pm$0.04 to 14.8$\pm$0.8 for OJ~248, and from 6.0$\pm$0.1 to 89.0$\pm$8.1 for 4C~+38.41.

Error budget analysis results showed that the systematically dominant parameter was the epoch selection, which has a significant impact on the distance estimates.
These results indicate that the uncertainties are dominated by systematic effects that significantly limit the reliability of the derived distances.

This study identifies the systematic limitations of the current methodology and delineates the key factors that constrain the precision of AGN variability–based distance estimates.
In particular, the strong sensitivity to observational and methodological choices indicates that the present framework is not yet robustly applicable to high-redshift blazars, leading to e.g., large deviations from the $\Lambda$CDM expectations.

These results suggest that the discrepancies are not solely due to statistical uncertainties, but arise from fundamental limitations in the current treatment of variability time-scales, flare structure, and the connection between causality-based sizes and physical emission regions.
In particular, the presence of overlapping flares and the lack of rigorous decomposition can lead to systematic overestimation of distances, especially at high redshift.
In addition, the scaling factor $f$, which links the variability time-scale to the physical size of the emitting region, remains poorly constrained and introduces an additional source of systematic uncertainty.

Therefore, substantial methodological and observational improvements as well as the underlying physical assumptions—including higher-cadence monitoring~(e.g., biweekly observations), robust flare decomposition, and better constraints on intrinsic source properties and scaling relations—are required before the cosmological applicability of this method can be reliably assessed.

%%% ACKNOWLEDGMENTS (IF ANY) %%%%%%%%%%%%%%%%%%%%%%%%%%%%%%%%%%%%%%%%

\acknowledgments

The Very Long Baseline Array~(VLBA) is an instrument of the National Radio Astronomy Observatory.
The National Radio Astronomy Observatory is a facility of the National Science Foundation operated by Associated Universities, Inc.
We acknowledge the use of data, analysis tools and services from the Open Universe platform.
This work was supported by a National Research Foundation of Korea~(NRF) grant funded by the Korean government~(MIST)~(2020R1A2C2009003, RS-2025-00562700).
J.A.H. acknowledges that this work was supported by the National Research Foundation of Korea~(NRF) grant funded by the Korea government~(MSIT) RS-2025-16302968.

%%% CALL LIST OF REFERENCES (natbib STYLE) %%%%%%%%%%%%%%%%%%%%%%%%%%
% \bibliography{jkas-sample}

% \bibliographystyle{aasjournal+jkas}
\bibliography{references}

%%% APPENDICES (IF ANY) %%%%%%%%%%%%%%%%%%%%%%%%%%%%%%%%%%%%%%%%%%%%%

\appendix
\section{Table of the circular Gaussian model-fitted values of the core measurements}
\label{Table_of_the_circular_Gaussian_model-fitted_values}
\begin{table*}[h]
    \centering
    \begin{threeparttable}
        \caption{OJ 248}
        \label{tabA1}
        \begin{tabular}{ccccc}
            \toprule[1.5pt]
            Flare & MJD & Flux Density (Jy) & Major FWHM (mas) & Beam (maj$\times$min, PA) (mas$\times$mas), ($^\circ$) \\
            No. &  & (1) & (2) & (3) \\ \hline
            1 & 55163 & 0.328$\pm$0.031 & 0.115$\pm$0.008 & 0.601$\times$0.280, 27.0 \\
             & 55206 & 0.576$\pm$0.051 & 0.047$\pm$0.003 & 0.373$\times$0.238, 11.1 \\
             & 55238 & 0.848$\pm$0.087 & 0.020$\pm$0.000 & 0.382$\times$0.232, 1.35 \\
             & 55261 & 1.117$\pm$0.079 & 0.044$\pm$0.002 & 0.420$\times$0.215, -5.13 \\
             & 55335 & 1.270$\pm$0.082 & 0.034$\pm$0.002 & 0.386$\times$0.230, -2.60 \\ \hline
            2 & 55361 & 0.855$\pm$0.049 & 0.038$\pm$0.002 & 0.379$\times$0.211, 2.8 \\
             & 55409 & 1.021$\pm$0.007 & 0.019$\pm$0.000 & 0.354$\times$0.207, -3.85 \\
             & 55429 & 1.484$\pm$0.110 & 0.021$\pm$0.001 & 0.359$\times$0.215, -4.61 \\
             & 55457 & 1.586$\pm$0.110 & 0.021$\pm$0.001 & 0.386$\times$0.214, -12.8 \\
             & 55493 & 2.789$\pm$0.159 & 0.025$\pm$0.001 & 0.385$\times$0.230, -4.77 \\
             & 55501 & 3.382$\pm$0.170 & 0.046$\pm$0.002 & 0.480$\times$0.203, -0.0526 \\
             & 55506 & 3.057$\pm$0.195 & 0.048$\pm$0.002 & 0.429$\times$0.210, -0.623 \\
             & 55513 & 2.851$\pm$0.180 & 0.055$\pm$0.002 & 0.414$\times$0.199, -3.34 \\
             & 55534 & 1.856$\pm$0.122 & 0.050$\pm$0.002 & 0.374$\times$0.223, -0.75 \\
             & 55563 & 2.080$\pm$0.117 & 0.041$\pm$0.002 & 0.401$\times$0.227, -5.97 \\
             & 55596 & 1.644$\pm$0.089 & 0.067$\pm$0.003 & 0.358$\times$0.210, -7.02 \\
             & 55621 & 1.397$\pm$0.136 & 0.075$\pm$0.005 & 0.394$\times$0.251, -6.71 \\
             & 55672 & 0.801$\pm$0.041 & 0.031$\pm$0.001 & 0.385$\times$0.212, 0.907 \\
             & 55703 & 1.009$\pm$0.050 & 0.049$\pm$0.002 & 0.397$\times$0.193, -12.5 \\
             & 55724 & 0.845$\pm$0.050 & 0.052$\pm$0.002 & 0.389$\times$0.223, 2.68 \\
             & 55763 & 0.537$\pm$0.047 & 0.045$\pm$0.003 & 0.378$\times$0.231, -1.57 \\
             & 55796 & 0.448$\pm$0.005 & 0.041$\pm$0.000 & 0.514$\times$0.353, -1.96 \\
             & 55820 & 0.359$\pm$0.007 & 0.028$\pm$0.000 & 0.389$\times$0.235, -5.28 \\ 
             & 56152 & 0.229$\pm$0.022 & 0.103$\pm$0.006 & 0.441$\times$0.243, -6.05 \\
             & 56207 & 0.465$\pm$0.021 & 0.065$\pm$0.002 & 0.438$\times$0.235, 8.95 \\
             & 56220 & 0.475$\pm$0.019 & 0.067$\pm$0.002 & 0.482$\times$0.229, 4.53 \\
             & 56227 & 0.433$\pm$0.042 & 0.040$\pm$0.003 & 0.460$\times$0.225, 19.3 \\
             & 56228 & 0.500$\pm$0.055 & 0.214$\pm$0.000 & 0.391$\times$0.224, 3.49 \\ \hline
            3 & 56282 & 0.476$\pm$0.018 & 0.032$\pm$0.001 & 0.400$\times$0.201, -0.802 \\
             & 56307 & 1.106$\pm$0.046 & 0.028$\pm$0.001 & 0.395$\times$0.229, -3.04 \\
             & 56349 & 1.855$\pm$0.062 & 0.025$\pm$0.001 & 0.400$\times$0.239, 9.59 \\
             & 56399 & 1.440$\pm$0.144 & 0.039$\pm$0.003 & 0.416$\times$0.202, -4.63 \\
             & 56442 & 1.155$\pm$0.067 & 0.043$\pm$0.002 & 0.412$\times$0.213, -6.06 \\
             & 56473 & 0.930$\pm$0.024 & 0.040$\pm$0.001 & 0.365$\times$0.208, -0.605 \\
             & 56501 & 1.085$\pm$0.032 & 0.048$\pm$0.001 & 0.337$\times$0.215, 3.73 \\
             & 56530 & 0.965$\pm$0.046 & 0.056$\pm$0.002 & 0.354$\times$0.211, 4.38 \\
             & 56614 & 0.521$\pm$0.039 & 0.052$\pm$0.003 & 0.341$\times$0.221, -13.9 \\
             & 56642 & 0.461$\pm$0.031 & 0.050$\pm$0.002 & 0.417$\times$0.240, -4.91 \\
             & 56677 & 0.340$\pm$0.012 & 0.064$\pm$0.002 & 0.363$\times$0.226, -4.09 \\
             & 56713 & 0.303$\pm$0.037 & 0.059$\pm$0.005 & 0.397$\times$0.217, -6.18 \\
            \bottomrule[1.5pt]
        \end{tabular}
        \begin{tablenotes}
	   \small
            \item{Note.~(1)~flux density of the emission region, (2)~the measured angular size of the emission region, (3)~beam size.}
        \end{tablenotes}
    \end{threeparttable}
\end{table*}

\begin{table*}[h]
    \centering
    \begin{threeparttable}
        \caption{4C +38.41}
        \label{tabA2}
        \begin{tabular}{ccccc}
            \toprule[1.5pt]
            Flare & MJD & Flux Density (Jy) & Major FWHM (mas) & Beam (maj$\times$min, PA) (mas$\times$mas), ($^\circ$) \\
            No. &  & (1) & (2) & (3) \\ \hline
            1 & 55534 & 2.244$\pm$0.246 & 0.061$\pm$0.005 & 0.338$\times$0.216, -14.2 \\
             & 55563 & 2.425$\pm$0.227 & 0.063$\pm$0.004 & 0.381$\times$0.228, -11.4 \\
             & 55596 & 2.752$\pm$0.178 & 0.067$\pm$0.003 & 0.324$\times$0.207, -18.3 \\
             & 55621 & 2.916$\pm$0.243 & 0.067$\pm$0.004 & 0.369$\times$0.233, -9.35 \\
             & 55672 & 3.038$\pm$0.111 & 0.039$\pm$0.001 & 0.333$\times$0.211, -13.3 \\
             & 55703 & 3.670$\pm$0.134 & 0.042$\pm$0.001 & 0.347$\times$0.206, -13.7 \\
             & 55724 & 3.519$\pm$0.170 & 0.036$\pm$0.001 & 0.340$\times$0.230, -11.4 \\ 
             & 55763 & 3.429$\pm$0.235 & 0.057$\pm$0.003 & 0.342$\times$0.229, -13.9 \\
             & 55796 & 3.006$\pm$0.147 & 0.059$\pm$0.002 & 0.420$\times$0.258, 10.7 \\
             & 55820 & 2.212$\pm$0.171 & 0.060$\pm$0.003 & 0.339$\times$0.227, -14.8 \\ \hline
            2 & 55991 & 1.370$\pm$0.050 & 0.054$\pm$0.001 & 0.366$\times$0.281, -6.37 \\
             & 56019 & 1.721$\pm$0.078 & 0.046$\pm$0.001 & 0.322$\times$0.204, -15.2 \\
             & 56073 & 2.099$\pm$0.230 & 0.043$\pm$0.003 & 0.367$\times$0.254, 18.6 \\
             & 56112 & 3.083$\pm$0.232 & 0.049$\pm$0.003 & 0.340$\times$0.221, -19.2 \\
             & 56152 & 3.160$\pm$0.256 & 0.043$\pm$0.002 & 0.378$\times$0.241, -5.86 \\
             & 56207 & 3.948$\pm$0.152 & 0.063$\pm$0.002 & 0.382$\times$0.232, -24.1 \\
             & 56220 & 3.746$\pm$0.242 & 0.091$\pm$0.004 & 0.414$\times$0.224, -20.7 \\
             & 56227 & 3.550$\pm$0.120 & 0.068$\pm$0.002 & 0.411$\times$0.242, -19.6 \\ 
             & 56228 & 3.817$\pm$0.206 & 0.075$\pm$0.003 & 0.348$\times$0.221, -18.3 \\ 
             & 56282 & 2.861$\pm$0.178 & 0.059$\pm$0.003 & 0.327$\times$0.207, -15.1 \\ \hline
            3 & 56307 & 2.352$\pm$0.046 & 0.028$\pm$0.000 & 0.341$\times$0.223, -11.2 \\
             & 56349 & 3.588$\pm$0.032 & 0.040$\pm$0.000 & 0.371$\times$0.234, 0.174 \\
             & 56399 & 4.596$\pm$0.081 & 0.043$\pm$0.001 & 0.388$\times$0.198, -12.4 \\
             & 56442 & 4.418$\pm$0.209 & 0.049$\pm$0.002 & 0.370$\times$0.236, -13 \\
             & 56473 & 3.692$\pm$0.239 & 0.043$\pm$0.002 & 0.335$\times$0.222, -13.8 \\
             & 56501 & 4.595$\pm$0.187 & 0.040$\pm$0.001 & 0.347$\times$0.214, -10.4 \\
             & 56530 & 4.618$\pm$0.213 & 0.042$\pm$0.001 & 0.339$\times$0.211, -16.3 \\
             & 56614 & 4.831$\pm$0.194 & 0.038$\pm$0.001 & 0.334$\times$0.210, -19.0 \\
             & 56642 & 4.356$\pm$0.177 & 0.055$\pm$0.002 & 0.342$\times$0.221, -5.87 \\
             & 56677 & 2.684$\pm$0.058 & 0.032$\pm$0.000 & 0.327$\times$0.216, -16.4 \\
             & 56713 & 2.537$\pm$0.162 & 0.047$\pm$0.002 & 0.335$\times$0.209, -16.3 \\
             & 56780 & 2.013$\pm$0.091 & 0.044$\pm$0.001 & 0.365$\times$0.216, -8.06 \\
             & 56828 & 1.902$\pm$0.082 & 0.048$\pm$0.001 & 0.351$\times$0.220, -11.3 \\
             & 56866 & 1.684$\pm$0.059 & 0.028$\pm$0.001 & 0.365$\times$0.205, -11.6 \\
             & 56923 & 1.497$\pm$0.059 & 0.043$\pm$0.001 & 0.380$\times$0.238, -11.7 \\
             & 56976 & 1.011$\pm$0.091 & 0.058$\pm$0.004 & 0.338$\times$0.230, -8.58 \\
             & 56996 & 0.990$\pm$0.038 & 0.053$\pm$0.001 & 0.318$\times$0.218, -13.2 \\
             & 57020 & 0.959$\pm$0.003 & 0.052$\pm$0.000 & 0.324$\times$0.203, -13.7 \\ \hline
            4 & 58110 & 1.430$\pm$0.166 & 0.071$\pm$0.006 & 0.484$\times$0.253, -16.2 \\
             & 58166 & 2.102$\pm$0.084 & 0.066$\pm$0.002 & 0.436$\times$0.231, -21.2 \\
             & 58187 & 2.484$\pm$0.031 & 0.069$\pm$0.001 & 0.417$\times$0.228, -27.3 \\ \hline
            5 & 58227 & 2.285$\pm$0.075 & 0.041$\pm$0.001 & 0.346$\times$0.211, -17.6 \\
             & 58249 & 3.608$\pm$0.225 & 0.072$\pm$0.003 & 0.320$\times$0.210, -15.7 \\
             & 58285 & 3.719$\pm$0.129 & 0.056$\pm$0.001 & 0.397$\times$0.220, -27.2 \\
             & 58315 & 3.083$\pm$0.202 & 0.053$\pm$0.002 & 0.337$\times$0.211, -3.33 \\
             & 58356 & 2.146$\pm$0.292 & 0.047$\pm$0.004 & 0.330$\times$0.210, -10.9 \\
             & 58406 & 2.936$\pm$0.191 & 0.015$\pm$0.001 & 0.352$\times$0.212, -10.9 \\
             & 58460 & 2.858$\pm$0.103 & 0.031$\pm$0.001 & 0.339$\times$0.211, -13.2 \\
             & 58493 & 2.230$\pm$0.092 & 0.097$\pm$0.003 & 0.334$\times$0.211, -12.0 \\
             & 58517 & 1.431$\pm$0.063 & 0.030$\pm$0.001 & 0.407$\times$0.238, 4.86 \\
             & 58522 & 1.297$\pm$0.217 & 0.084$\pm$0.010 & 0.340$\times$0.213, -11.4 \\
            \bottomrule[1.5pt]
        \end{tabular}
        \begin{tablenotes}
	   \small
            \item{Note.~(1)~flux density of the emission region, (2)~the measured angular size of the emission region, (3)~beam size.}
        \end{tablenotes}
    \end{threeparttable}
\end{table*}

\section{Estimated distances from each source}
\label{Estimated_distances_from_each_source}
\begin{table*}[h]
    \centering
    \begin{threeparttable}
        \caption{OJ~248}
        \label{tabB1}
        \begin{tabular}{cccccccc}
            \toprule[1.5pt]
            Flare & Epoch & \multicolumn{2}{c}{$D^{\text{eq}}_{\text{A}}$~(Mpc)~(1)} & $D^{\text{Cheong}}_{\text{A}}$~(Mpc) & $D^{{\delta_{\text{IC}}}}_{\text{A}}$~(Mpc) \\
            No. & (MJD) & Disk & Sphere & (2) & (3) \\ \hline
            1 & 55163 & 182.7$\pm$41.9 & 193.1$\pm$44.3 & 15.7-8506.6 & 17.1$\pm$3.8 \\
            & 55206 & 2692.6$\pm$578.6 & 2846.4$\pm$611.6 & 236.5-123907.8 & 333.0$\pm$68.4 \\
            & 55238 & 33829.0$\pm$8160.6 & 35762.1$\pm$8626.9 & 2871.2-1590514.1 & 4443.0$\pm$971.6 \\
            & 55261 & 3191.1$\pm$586.0 & 3373.4$\pm$619.5 & 291.3-143071.8 & 794.5$\pm$138.2 \\
            & 55335 & 7101.5$\pm$1232.3 & 7507.3$\pm$1302.7 & 656.6-315675.5 & \\ \hline
            2 & 55361 & 8729.2$\pm$1190.2 & 9228.0$\pm$1258.2 & 843.3-375734.7 & \\
            & 55409 & 67295.8$\pm$4205.4 & 71141.3$\pm$4445.8 & 7057.1-2708381.1 & \\
            & 55429 & 47704.2$\pm$7629.6 & 50430.2$\pm$8065.6 & 4482.6-2095979.0 & \\
            & 55457 & 51477.8$\pm$7521.7 & 54419.4$\pm$7951.5 & 4916.7-2234829.6 & \\
            & 55493 & 29181.2$\pm$3838.7 & 30848.7$\pm$4058.1 & 2834.7-1250752.9 & \\
            & 55501 & 4962.1$\pm$601.6 & 5245.7$\pm$636.0 & 487.7-210748.9 & 1132.6$\pm$126.2 \\ \hline
            3 & 56282 & 4682.5$\pm$457.5 & 4950.1$\pm$483.7 & 472.6-194697.8 & 173.3$\pm$15.9 \\
            & 56307 & 6993.7$\pm$544.3 & 7393.4$\pm$575.4 & 721.4-285532.0 & 568.8$\pm$19.6 \\
            & 56349 & 10081.6$\pm$889.8 & 10657.7$\pm$940.7 & 1028.2-415583.7 & 1278.7$\pm$102.8 \\
            \bottomrule[1.5pt]
        \end{tabular}
        \begin{tablenotes}
    	  \small
            \item{Note.~(1)~Estimated $D_{\text{A}}$ using $T_{\text{eq}}$~(Mpc), (2)~estimated $D_{\text{A}}$ using $T^{\text{Cheong}}_{\text{B,~int}}$, (3)~estimated $D_{\text{A}}$ using $\delta_{\text{IC}}$.}
        \end{tablenotes}
    \end{threeparttable}
\end{table*}

\begin{table*}[h]
    \centering
    \begin{threeparttable}
        \caption{4C~+38.41}
        \label{tabB2}
            \begin{tabular}{ccccccccc}
            \toprule[1.5pt]
                Flare & Epoch & \multicolumn{2}{c}{$D^{\text{eq}}_{\text{A}}$~(Mpc)~(1)} & $D^{\text{Cheong}}_{\text{A}}$~(Mpc) & $D^{\delta_{\text{IC}}}_{\text{A}}$~(Mpc) \\
                No. & (MJD) & Disk & Sphere & (2) & (3)\\ \hline
                1 & 55534 & 9405.2$\pm$2742.9 & 9942.6$\pm$2899.7 & 745.2-460156.2 & \\
                 & 55563 & 8746.9$\pm$2314.0 & 9246.8$\pm$2446.2 & 719.6-418972.9 & \\
                 & 55596 & 7194.8$\pm$1597.3 & 7605.9$\pm$1688.6 & 626.1-333036.0 & \\
                 & 55621 & 7344.7$\pm$1828.7 & 7764.4$\pm$1933.2 & 617.0-347477.5 & \\
                 & 55672 & 37608.8$\pm$7034.1 & 39757.9$\pm$7436.1 & 3420.0-1691022.2 & 7729.1$\pm$1437.1 \\
                 & 55703 & 29944.2$\pm$5622.6 & 31655.3$\pm$5943.9 & 2720.6-1347226.4 & 6860.1$\pm$120.6 \\ \hline
                2 & 55991 & 8218.6$\pm$815.1 & 8688.2$\pm$861.7 & 828.1-342184.7 & \\
                 & 56019 & 13381.8$\pm$1529.8 & 14146.5$\pm$1617.2 & 1325.7-564834.4 & \\
                 & 56073 & 16597.3$\pm$4037.6 & 17545.8$\pm$4268.3 & 1404.9-781626.4 & 2492.6$\pm$605.1 \\
                 & 56112 & 11358.4$\pm$1970.1 & 12007.5$\pm$2082.6 & 1050.1-504867.2 & 2559.4$\pm$435.3 \\
                 & 56152 & 16563.1$\pm$3046.7 & 17509.6$\pm$3220.8 & 1511.9-742796.7 & \\
                 & 56207 & 5241.8$\pm$539.9 & 5541.3$\pm$570.7 & 525.9-219002.7 & \\ \hline
                3 & 56307 & 46967.6$\pm$2714.2 & 49651.5$\pm$2869.3 & 4950.0-1881885.7 & \\
                 & 56349 & 16179.0$\pm$773.9 & 17103.5$\pm$818.1 & 949.1-642155.7 & \\
                 & 56399 & 12963.9$\pm$746.0 & 13704.7$\pm$788.7 & 1366.6-519317.3 & 3024.0$\pm$178.5 \\ \hline
                4 & 58110 & 1487.8$\pm$447.8 & 1572.8$\pm$473.4 & 116.3-73318.6 & \\
                 & 58166 & 1883.2$\pm$351.9 & 1990.9$\pm$372.0 & 171.3-84664.7 & \\
                 & 58187 & 1584.5$\pm$265.3 & 1675.0$\pm$280.4 & 147.5-70067.2 & 420.6$\pm$70.5 \\ \hline
                5 & 58227 & 10736.8$\pm$1313.6 & 11350.3$\pm$1388.6 & 1054.1-456453.6 & \\
                 & 58249 & 1940.0$\pm$337.9 & 2050.8$\pm$357.3 & 179.2-86284.9 & \\
                 & 58285 & 4077.2$\pm$525.1 & 4310.1$\pm$555.1 & 397.4-174326.9 & \\
            \bottomrule[1.5pt]
        \end{tabular}
        \begin{tablenotes}
	   \small
            \item{Note.~(1)~Estimated $D_{\text{A}}$ using $T_{\text{eq}}$~(Mpc), (2)~estimated $D_{\text{A}}$ using $T^{\text{Cheong}}_{\text{B,~int}}$, (3)~estimated $D_{\text{A}}$ using $\delta_{\text{IC}}$.}
        \end{tablenotes}
    \end{threeparttable}    
\end{table*}

%%% THE END %%%

\end{document}